\documentclass[pra,aps,superscriptaddress,secnumarabic,eqsecnum,showpacs]{revtex4}
\usepackage{epsfig,amsmath}
\begin{document}

\title{Quantal Two-Centre Coulomb Problem treated by means of the 
Phase-Integral Method I. General Theory}

\author{N. Athavan} 

\thanks{Present address: Department of Physics, Government Arts
College, Ariyalur - 621 713, India.}

\affiliation{Centre for Nonlinear Dynamics, Department of Physics,
Bharathidasan University, Tiruchirapalli 620 024,India}

\author{N. Fr\"oman} 

\affiliation{Department of Theoretical Physics, University of Uppsala,
Box 803, S-751 05 Uppsala, Sweden}

\author{P.O. Fr\"oman} 

\affiliation{Department of Theoretical Physics, University of Uppsala,
Box 803, S-751 05 Uppsala, Sweden}

\author{M. Lakshmanan} 

\affiliation{Centre for Nonlinear Dynamics, Department of Physics, 
Bharathidasan University, Tiruchirapalli 620 024,India}

\begin{abstract}

The present paper concerns the derivation of phase-integral
quantization conditions for the two-centre Coulomb problem under the
assumption that the two Coulomb centres are fixed. With this
restriction we treat the general two-centre Coulomb problem according
to the phase-integral method, in which one uses an {\it a priori}
unspecified {\it base function}. We consider base functions containing
three unspecified parameters $C, \tilde C$ and $\Lambda$. When the
absolute value of the magnetic quantum number $m$ is not too small, it
is most appropriate to choose $\Lambda=|m|\ne 0$. When, on the other
hand, $|m|$ is sufficiently small, it is most appropriate to choose
$\Lambda = 0$.  Arbitrary-order phase-integral quantization conditions
are obtained for these choices of $\Lambda$. The parameters $C$ and
$\tilde C$ are determined from the requirement that the results of the
first and the third order of the phase-integral approximation coincide,
which makes the first-order approximation as good as possible.

In order to make the paper to some extent self-contained, a short
review of the phase-integral method is given in the Appendix.

\end{abstract}
\maketitle

\section{Introduction}
\label{sec1}
The two-centre Coulomb problem, that is, the problem of solving the
Schr\"odinger equation for the motion of an electron with the charge
$-e$ $(e>0)$ in the field of two fixed Coulomb centres with charges $Z_1e$
and $Z_2e$ at the distance $r_{12}$ from each other, plays an important
role in several fields of theoretical physics, for example in the
theory of diatomic molecules, in scattering theory, and in the
three-body problem. The two-centre Coulomb problem has therefore been
the subject of extensive studies both by numerical and by approximate
analytical methods, and hence the literature on this problem is very
comprehensive. In spite of this fact it is still of interest to continue
the treatment of this problem for arbitrary values of $Z_1, Z_2$ and
$r_{12}$. One reason for this is that the numerically exact solution of
the problem meets with difficulties when $|Z_1-Z_2|$ increases. There
appear also numerical difficulties for large values of $r_{12}$. For a
general review of the problem we refer to Eyring {\it et al.} \cite{ery}, 
Herzberg \cite{her}, Slater \cite{slate} and Rosen \cite{rose}.

Ignoring the finiteness of the mass of the protons, Bates {\it et
al.}\cite{bat} obtained important numerical results for the hydrogen
molecule ion. Corresponding numerical results were obtained by Wallis
and Hulburt \cite{wall} for the homonuclear one-electron two-centre
problem, by Bates and Carson \cite{bat2} for the ion
$HeH^{2+}(Z_1=1,Z_2=2)$, by Wind \cite{win} for the ground state of the
hydrogen molecule ion, by Peek \cite{peek} for the states $1s\sigma_g$ and
$2p\sigma_u$ of the hydrogen molecule ion, and by Ponomarev and Puzynina 
\cite{pon,pon2} for several states of the Coulomb two-centre system 
with $Z_1=1$ and $Z_2=2,3,...,8$. Hunter and Pritchard\cite{hun} used a numerical
procedure to compute nonadiabatic energies for the first few
rotation-vibration levels of $^2\Sigma_gH_2^+$, $^2\Sigma HD^+$ and
$^2\Sigma_gD_2^+$. For the hydrogen molecule ion Rosenthal and E. Bright
Wilson, Jr., \cite{wil} calculated an accurate value of the ground
state energy which is in agreement with the values obtained by Wind \cite{win}
and Peek\cite{peek}. For different internuclear distances Bates and Reid
\cite{bat3}, Murai \cite{mur} and Murai and Takatsu \cite{mur2} 
calculated electronic energies for various states of the hydrogen
molecule ion under the assumption of infinite proton mass. For a large
range of internuclear distances Winter {\it et al.}\cite{wint} made
very accurate calculations for the lowest 20 states of the molecule ion
$HeH^{2+}$. Klaus\cite{kla} studied the electronic energy of the ground
state of the hydrogen molecule ion for small internuclear separation
$r_{12}$ and confirmed the remarkable discovery by Byers Brown and Steiner \cite{bye}
that the electronic energy cannot be expanded in powers of $r_{12}$
alone, but that powers of ln$r_{12}$ must also be included. Klaus \cite{kla} also
obtained for the hydrogen molecule ion further terms in the series given 
by Byers Brown and Steiner \cite{bye} for the general two-centre Coulomb problem.

There exist also various approximate analytical methods for solving the
quantal two-centre Coulomb problem. In one of them one uses the
quasiclassical approximation, that is, the first order of the
phase-integral approximation. In using this method, one encountered in
the early papers difficulties associated with the divergence of the
phase-integral due to an inappropriate choice of the
phase-integrand (also called the quasimomentum), which is not determined
quite uniquely. Different authors have mastered these difficulties in
different ways, but no single unifying method has until now been
proposed for the two-centre Coulomb problem. An essential feature
in our method is the use of the phase-integral approximation generated
from an unspecified base function; see the Appendix.

The semiclassical quantization of the low-lying electronic states of the 
hydrogen molecule ion was treated by Strand and Reinhardt \cite{strand}. 
Pajunen \cite{paj} calculated the energy levels of the hydrogen molecule
ion (under the assumption of infinitely heavy nuclei) in the first and
the third order of the phase-integral approximation. 
Although he refers to one of the papers (his reference 10) in which the
phase-integral approximation generated from an unspecified base function
was introduced, he does not use the freedom to choose optimal
expressions for the
functions that he denotes by $Q_{mod}(\mu)$ and $Q_{mod}(\nu)$, and
that in the present paper correspond to the more general base functions
$\tilde Q(\xi)$ and $Q(\eta)$, respectively. In the present paper we
shall make full use of the possibility to choose $\tilde Q(\xi)$ and
$Q(\eta)$ most conveniently.

The phase-integral method, in which one uses the phase-integral
approximation generated from an unspecified base function \cite{fro10},
 offers a method for mastering the previously
mentioned difficulties in a unified way for an arbitrary order of
the phase-integral approximation. In the present paper we shall apply
this method to the quantal two-centre Coulomb problem with fixed Coulomb
centres.

For the convenience of the reader a short review of the phase-integral
method and formulas to be used are given in the Appendix.

The present paper will be the basis for further work, in which
convenient transformations to complete elliptic integrals will be used,
analogously as in papers by Lakshmanan and Kaliappan\cite{lak} and
Lakshmanan {\it et al.}\cite{lak2,lak3}.
Thereby the contour integrals, occurring in the quantization conditions,
will be expressed in terms of complete elliptic integrals.

\section{Separation of the Schr\"odinger equation in elliptic coordinates}
\label{sec2}
We start by quoting some well-known results.
The time-independent Schr\"odinger equation for the motion of an electron of 
mass $\mu$ and charge $-e$ $(e>0)$ in the field of two fixed Coulomb centres 
with charges $Z_1e$ and $Z_2e$ takes the following form

\begin{equation}
\label{e1}
\left (-\frac{\hbar ^2}{2\mu}\Delta_{\vec
r}-\frac{Z_1e^2}{r_1}-\frac{Z_2e^2}{r_2}\right ) \Psi (\vec r) = E\Psi(\vec
r),
\end{equation}
where $r_1$ and $r_2$ are the distances of the electron from the two
centres, $\vec r$ is the position vector of the electron, and E is the
electronic energy. To obtain the total energy one must add the potential
energy of the two fixed charges, getting
\begin{equation}
E_{total}=\frac{Z_1Z_2e^2}{r_{12}}+E,
\end{equation}
where $r_{12}$ is the distance between the two centres. The differential
equation (\ref{e1}) is separable in elliptic coordinates. If one introduces 
the variables 
\begin{subequations}
\begin{equation}
\xi=\frac{r_1+r_2}{r_{12}},\hskip 10pt 1\le \xi < +\infty,
\end{equation}
\begin{equation}
\eta=\frac{r_1-r_2}{r_{12}},\hskip 10pt -1\le \eta \le +1,
\end{equation}
\end{subequations}
and puts
\begin{equation}
\label{wf}
\Psi (\vec r)=X(\xi)Y(\eta)e^{im\phi},
\end{equation}
where $m$ is the magnetic quantum number 
(positive or negative integer or zero), and $\phi$ is the corresponding angle, 
the separation yields, in atomic units $(\hbar =e=\mu=1)$, the two 
differential equations
\begin{subequations}
\label{e2}
\begin{equation}
\frac{d}{d\xi}\left((\xi^2-1)\frac{dX}{d\xi}\right)+\left(-p^2\xi^2+b'
\xi+A-\frac{m^2}{\xi^2-1}\right)X = 0,
\end{equation}
\begin{equation}
\frac{d}{d\eta}\left((1-\eta^2)\frac{dY}{d\eta}\right)+\left(p^2\eta^2+b
\eta-A-\frac{m^2}{1-\eta^2}\right)Y = 0,
\end{equation}
\end{subequations}
where $A$ is the separation constant and
\begin{subequations}
\begin{equation}
p^2=-\frac{1}{2}r_{12}^2E, 
\end{equation}
\begin{equation}
b'=r_{12}(Z_2+Z_1), 
\end{equation}
\begin{equation}
b=r_{12}(Z_2-Z_1).
\end{equation}
\end{subequations}
Putting
\begin{subequations}
\begin{equation}
X(\xi)=\frac{f(\xi)}{(\xi^2-1)^\frac{1}{2}},
\end{equation}
\begin{equation}
Y(\eta)=\frac{g(\eta)}{(1-\eta^2)^\frac{1}{2}},
\end{equation}
\end{subequations}
we can transform the differential equations (\ref{e2}) into
\begin{subequations}
\label{e3}
\begin{equation}
\left(\frac{d^2}{d\xi^2}+\tilde R(\xi)\right)f(\xi)=0,
\end{equation}
\begin{equation}
\left(\frac{d^2}{d\eta^2}+R(\eta)\right)g(\eta)=0,
\end{equation}
\end{subequations}
where
\begin{subequations}
\label{e4}
\begin{equation}
\tilde R(\xi)= -p^2+\frac{b'\xi+A'}{\xi^2-1}-\frac{m^2-1}{(\xi^2-1)^2},
\end{equation}
\begin{equation}
R(\eta)= -p^2+\frac{b\eta-A'}{1-\eta^2}-\frac{m^2-1}{(1-\eta^2)^2},
\end{equation}
\end{subequations}
with
\begin{equation}
A'=A-p^2.
\end{equation}
The differential equations (\ref{e3}) are of the Schr\"odinger type. By
solving them simultaneously under the boundary conditions that
$f(+1)=f(+\infty)=0$ and $g(-1)=g(+1)=0$, one can obtain the energy and
the separation constant as functions of the distance $r_{12}$ and the
quantum numbers.

When $Z_1=Z_2$ every bound-state wave function $Y(\eta)$ is either an
even or an odd function of $\eta$, and when $\vec r$ is reflected at the
centre of symmetry for the two-centre Coulomb problem, the wave function
(\ref{wf}) remains unchanged when $Y(\eta)$ and $m$ are both even or
both odd, while the wave function (\ref{wf}) changes sign when one of
$Y(\eta)$ and $m$ is even and the other is odd.

\section{Application of the phase-integral method}
\label{sec3}
The essential features of the phase-integral method are briefly sketched
in the Appendix. The phase-integral solutions of the differential
equations (2.8a) and (2.8b), respectively, are linear combinations of the
phase-integral functions
\begin{subequations}
\begin{equation}
\tilde q^{-\frac{1}{2}}(\xi)\mbox{exp}\left\{\pm i\int^\xi\tilde
q(\xi)d\xi\right\}
\end{equation}
and
\begin{equation}
q^{-\frac{1}{2}}(\eta)\mbox{exp}\left\{\pm i\int^\eta
q(\eta)d\eta\right\},
\end{equation}
\end{subequations}
respectively, where $\tilde q(\xi)$ and $q(\eta)$, respectively, are generated
according to (A5a,b), (A6a,b,c), (A3) and (A2) in the Appendix, with
$R(z), Q(z)$ replaced by the appropriate functions $\tilde R(\xi), \tilde
Q(\xi)$ and $R(\eta),Q(\eta)$, respectively, the quantities pertaining
to the $\xi$-equation being characterized by a tilde.

\subsection{Base functions generating phase-integral solutions}

As is seen from (2.9a,b), the functions $\tilde R(\xi)$ and $R(\eta)$
have poles at $\xi=\pm 1$ and $\eta =\pm1$, respectively; these poles
are of the second order if $m\ne \pm 1$, but of the first order if
$m=\pm 1$. Furthermore, we note that when $m\ne 0$ the coefficients of
the second-order poles differ from $\frac{1}{4}$, while for $m = 0$
they are equal to $\frac{1}{4}$.

There are two main alternatives, discussed in the Appendix for the case
of the radial Schr\"odinger equation, for choosing the base functions
generating the phase-integral functions (3.1a) and (3.1b). Unifying
these two alternatives, we choose the squares of the base functions to
be
\begin{subequations}
\begin{equation}
\tilde Q^2(\xi)=-p^2+\frac{A'-\tilde
C+b'\xi}{\xi^2-1}-\frac{\Lambda^2}{(\xi^2-1)^2},
\end{equation}
\begin{equation}
Q^2(\eta)=-p^2+\frac{-A'+
C+b\eta}{1-\eta^2}-\frac{\Lambda^2}{(1-\eta^2)^2},
\end{equation}
\end{subequations}
where $C, \tilde C$ and $\Lambda$ are parameters, the choice of which
we shall discuss below. The introduction of these parameters increases
essentially the flexibility of the phase-integral formulas to be
obtained.

By choosing $C=\tilde C=1/4$ one obtains in the limit when
$r_{12}\rightarrow 0$ the energy E and the reduced separation constant
$A'$ correctly from the first-order phase-integral quantization
conditions (to be derived in Section 3.2). For arbitrary values of
$r_{12}$ it is most appropriate to determine $C$ and $\tilde C$ as
functions of $r_{12}$ such that one obtains the same value of $p^2$
(i.e., of the energy) and also of $A'$ in the first and in the third
order of the phase-integral approximation. One thereby achieves the
optimal accuracy obtainable from the first-order quantization
conditions. We emphasize that this can be achieved by the use of the
phase-integral approximation generated from an unspecified base
function (described in the Appendix), but cannot be achieved by means
of the JWKB approximation. The decisive properties of the
phase-integral approximation in question versus the JWKB approximation
have been explained in some detail by Dammert and P.O. Fr\"{o}man
\cite{dam} and by Fr\"oman and Fr\"oman \cite{fro10}.

When $|m|$ is not too small, we choose $\Lambda=|m| \ne 0$, but when
$|m|$ is sufficiently small we choose $\Lambda=0$. For $m=0$ one should
always choose $\Lambda =0$. The two alternatives $\Lambda=|m| \ne 0$ and
$\Lambda=0$ yield solutions with different properties. When $\Lambda=|m| \ne 0$
the phase integral solutions of the differential equations (2.8a) and (2.8b)
remain valid as $\xi\rightarrow \pm 1$ and $\eta \rightarrow \pm
1$, respectively. When $\Lambda=0$ the phase-integral solutions of the
differential equations in question break down as $\xi\rightarrow
\pm 1$ and $\eta \rightarrow \pm 1$, respectively, but the regular
solutions of the $\xi$- and $\eta$-equations can be obtained at some
distances from those points by the use of the connection formula
described in subsection A.2.b of the Appendix, when one there replaces $l$
by $(|m|-1)/2$; cf. (2.9a,b). The wave functions obtained in that way
are expected to be the more accurate the stronger the Coulomb
singularities of $\tilde R(\xi)$ and $R(\eta)$ at $\xi = \pm 1$ and
$\eta = \pm 1$ are.  However, even if the Coulomb singularities are strong,
these wave functions are not expected to be good if
$|m|$ is too large, in which case one should use $\Lambda = |m|\ne 0$,
as mentioned in the beginning of this paragraph.

Using the terminology classically allowed region and classically
forbidden region in a generalized sense, {\it viz.} to characterize
regions where $\tilde Q^2(\xi)$ or $Q^2(\eta)$ is larger than zero and
less than zero, respectively, we shall now discuss the wave
functions pertaining to the $\xi$-equation and the $\eta$-equation. 

\subsubsection{Wave functions pertaining to the $\xi$-equation}

According to (3.2a) the
function $-\tilde Q^2(\xi)$, where $1\le \xi < +\infty$, always
corresponds to a single-well potential. 

For $\Lambda=|m|\ne 0$ there are four zeros of $\tilde Q^2(\xi)$, which
we denote by $\xi_1,\xi_2,\xi_3$ and $\xi_4$; see Fig. 1(a). The zeros
$\xi_1$ and $\xi_2$ may be real and both less than 1, or they may be
complex conjugate. The zeros $\xi_3(> 1)$ and $\xi_4(> \xi_3)$ are
real. There is thus a classically allowed region between $\xi_3$ and
$\xi_4$, but classically forbidden regions for $1< \xi < \xi_3$ and
$\xi > \xi_4$. The phase-integral wave functions generated from $\tilde
Q(\xi)$ are good at $\xi=1$, and we can use the arbitrary-order
connection formula, given by (A13) and (A14) in the Appendix, for
tracing the physically acceptable wave function from the classically
forbidden region between $1$ and $\xi_3$ to the classically allowed
region, that is, the region between $\xi_3$ and $\xi_4(> \xi_3)$ where
$\tilde Q^2(\xi)> 0$; see Fig. 1(a).

For $\Lambda =0$ there are two zeros of
$\tilde Q^2(\xi)$, which we denote by $\xi_3(< \xi_4)$ and $\xi_4(>
1)$. The classically allowed region lies between $\xi_3$ and $\xi_4$
when $\xi_3>1$, but between 1 and $\xi_4$ when $\xi_3<1$; see Fig. 1.
The phase-integral wave function is not good at $\xi =1$, but for the
physically acceptable wave function one can obtain a phase-integral
expression in the interior of the classically allowed region, when
$\xi_3>1$ and the classically forbidden region between $1$ and $\xi_3$
is sufficiently large, by means of the arbitrary-order connection
formula given by (A13) and (A14) in the Appendix, and when $\xi_3<1$ by
means of the connection formula presented in subsection A.2.b of the
Appendix.

Both when $\Lambda =|m|\ne 0$ and when $\Lambda =0$ the wave function,
obtained in the classically allowed region to the left of $\xi_4$ as
described above, can be joined to the physically acceptable wave function
traced from the classically forbidden region to the right of $\xi_4$
into the classically allowed region to the left of $\xi_4$ with the aid
of the arbitrary-order connection formula given by (A13) and (A14) in
the Appendix. In this way  alternative quantization conditions,
corresponding to $\Lambda=|m| \ne 0$ and $\Lambda=0$, can be obtained.
They can be combined into one quantization condition. Fig. 1 illustrates
the two possible situations that the classically allowed region is
delimited either by $\xi_3$ and $\xi_4$ [Fig. 1(a)] or by the pole at
$\xi=+1$ and the turning point at $\xi=\xi_4$ [Fig. 1(b)].

\subsubsection{Wave functions pertaining to the $\eta$-equation}
The function $-Q^2(\eta)$ may correspond either to a
single-well potential or to a double-well potential. 

For $\Lambda=|m|\ne 0$ the
phase-integral solution is valid at the poles $\eta = \pm 1$ (which
delimit classically forbidden regions) and can be traced into the classically
allowed region closest to the pole in question with the aid of the
connection formula given by (A13) and (A14) in the Appendix. When 
there is only one classically allowed region,
one obtains the quantization condition by identifying the two expressions for
the wave function in that region. This case applies when $-Q^2(\eta)$
is a single-well potential (Fig. 2), or when the energy lies so far above
the top of an underdense barrier [Fig. 4(b)] that it is appropriate
to disregard the presence of the complex conjugate zeros $\eta_2$ and $\eta_3$
of $Q^2(\eta)$. When $-Q^2(\eta)$ corresponds to 
a double-well potential, the wave function
can be traced from the region on one side of the barrier to the region
on the other side with the aid of the arbitrary-order
connection formula for a barrier described in Section A.3 of the
Appendix; see Fig. 3 and Fig. 4. Joining the two expressions for 
the wave function thus obtained to each other, one obtains the quantization 
condition.

When $\Lambda=0$ [Fig. 2 or Fig. 4(b)] the phase-integral wave 
function is not good at
$\eta=\pm 1$, but at some distance from these points physically
acceptable solutions can be obtained in the same classically allowed region  by
the use of the connection formulas presented in Sections A.2 and A.3 of
the Appendix. Thus one obtains two expressions for the wave function, and
by identifying these expressions one obtains a quantization condition.

\subsection{Quantization conditions}
\subsubsection{Quantization conditions pertaining to the $\xi$-equation}

For the
differential equation (2.8a) the physically relevant interval 
is $1 < \xi < \infty$. The phase-integral quantization
condition for the situation in Fig. 1(a) involves a contour integral in the
complex $\xi$-plane encircling $\xi_3$ and $\xi_4$, while the
phase-integral quantization condition for the situation in Fig. 1(b) involves
a contour integral encircling the simple pole at $\xi =1$ and the
generalized classical turning point $\xi_4$. The quantization condition
(A31) in the Appendix applies to the first situation, and the
quantization condition (A32) in the Appendix applies to the second
situation. Introducing the notations
\begin{subequations}
\begin{equation}
\tilde L=\sum_{n=0}^N \tilde L ^{(2n+1)},
\end{equation}
\begin{equation}
\tilde L^{(2n+1)}= \frac{1}{2}\int_{\Lambda_{\tilde L}} \tilde
q^{(2n+1)}(\xi)d\xi,
\end{equation}
\end{subequations}
\begin{subequations}
\begin{equation}
\tilde L'=\sum_{n=0}^N \tilde L^{'(2n+1)},
\end{equation}
\begin{equation}
\tilde L^{'(2n+1)} = \frac{1}{2}\int_{\Lambda_{\tilde L'}} \tilde
q^{(2n+1)}(\xi)d\xi,
\end{equation}
\end{subequations}
where $\tilde q(\xi)$ is obtained according to (A5a,b), (A6a,b,c), (A3)
and (A2) in the Appendix, and $\Lambda_{\tilde L}$ and $
\Lambda_{\tilde L'}$ are the appropriate contours of integration pertaining to
$\xi_3>1$ and $\xi_3<1$, respectively, and shown in Fig. 1(a) and Fig.
1(b), respectively, we can write the two quantization conditions
corresponding to $\xi_3>1$ and $\xi_3<1$ as follows
\begin{subequations}
\begin{equation}
\tilde L=\left (\tilde s + \frac{1}{2}\right)\pi,\hskip 10pt \xi_3>1,
\end{equation}
\begin{equation}
\tilde L'=\left (\frac{|m|}{2}+\tilde s + \frac{1}{2}\right)\pi,\hskip 10pt \xi_3<1,
\end{equation}
\end{subequations}
where $\tilde s$ is an integer. If, when $\Lambda=|m|\ne 0$ and hence
 $\xi_3>1$, we enlarge the
contour of integration $\Lambda_{\tilde L}$ in Fig. 1(a), so that the new
contour $\Lambda_{\tilde L'}$ encloses the turning points $\xi_3$ and
$\xi_4$ as well as the pole at $\xi =1$, and if we compensate in (3.5a)
along with (3.3a,b) for this change by taking the residue of the
integrand at $\xi=1$ into account, we obtain a general quantization
condition, valid for both cases $\xi_3>1$ and $\xi_3<1$, i.e., for both
situations depicted in Fig. 1, $viz$.
\begin{equation}
\tilde L'=\left (\frac{|m|}{2}+\tilde s + \frac{1}{2}\right)\pi,\hskip 10pt \hskip
5pt \xi_3>1 \mbox{ or } \xi_3<1.
\end{equation}
Besides condensing the two alternative quantization conditions (3.5a)
and (3.5b) nicely into one formula, the quantization condition (3.6) has
the further merit that, if the integration along the contour is made
numerically (in cases where expressions in terms of complete elliptic
integrals are not available), it may be advantageous to use the contour
$\Lambda_{\tilde L'}$ instead of $\Lambda_{\tilde L}$ when $\xi_3>1$.

The quantization condition (3.6) yields the value of the reduced separation
constant $A'$ as a function of $p^2$ and $\tilde C$; see (3.2a).

\subsubsection{Quantization conditions pertaining to the
$\eta$-equation}
In the physically relevant interval $-1<\eta<1$ the function
$-Q^2(\eta)$ may correspond to a single-well potential (Fig. 2) or to a
double-well potential with a superdense (Fig. 3) or underdense (Fig. 4)
barrier. When, in the case of an underdense barrier, the energy lies
sufficiently far above the top of the barrier, it may be preferable to
disregard the barrier and to treat the double-well potential problem as
a single-well potential problem.

When $\Lambda=|m|\ne 0$ or $\Lambda=0$, and $-Q^2(\eta)$ is or can be considered
as a single-well potential, and the classically allowed region is
delimited by two simple zeros of $Q^2(\eta)$, as shown in Fig. 2 and
in Fig. 4(a), we obtain from (A31) the single-well quantization
condition

\begin{equation}
\label{l1}
L=\left (s+\frac{1}{2}\right )\pi, \hskip 10pt s=\mbox{non-negative integer},
\end{equation}
where by definition
\begin{equation}
\label{l2}
L=\frac{1}{2}\int_{\Lambda_L}q(\eta)d\eta,
\end{equation}
$\Lambda_L$ being a closed contour encircling  the
generalized classical turning points. Note that in the derivation of (3.7) 
we have considered the classically forbidden regions to be thick also when
$\Lambda=0$ (Fig. 2). When $\Lambda=|m|\ne 0$ we can with the aid of residue calculus
write (\ref{l1}) along with (\ref{l2}) as 

\begin{equation}
L'=\left (|m|+s+\frac{1}{2}\right )\pi, 
\end{equation}
where $L'$ is defined by
\begin{equation}
L'=\frac{1}{2}\int_{\Lambda_{L'}}q(\eta)d\eta,
\end{equation}
$\Lambda_{L'}$ being a closed contour encircling $-1$ and $+1$;
see Fig. 2 and Fig. 4(a).

When $\Lambda=0$,  and the classically allowed region is delimited by two
first-order poles of $Q^2(\eta)$, as shown in Fig. 4(b), and the 
energy lies far above the top of the barrier, one can consider 
$-Q^2(\eta)$ as a single-well potential.
From (A33) one then obtains  the quantization condition (3.9) with $L'$
defined by (3.10), where $\Lambda_{L'}$ is now the contour 
in Fig. 4(b).

We disregard the possibility that $\Lambda =0$ and the residues of 
$Q^2(\eta)$ at $\eta =-1$ and $\eta =+1$ have different signs, since
this case has so far not appeared in the applications. 

 When $-Q^2(\eta)$ corresponds to a
double-well potential (Fig. 3 and Fig. 4), which is usually the case,
the quantization condition (A39) gives 
\begin{equation}
\mbox{cos}(\alpha+\beta+\tilde \phi -2a)={{\mbox{cos}(\alpha-\beta)}\over
{[1+\mbox{exp}(-2\pi\bar K)]^\frac{1}{2}}},
\end{equation}
where
\begin{subequations}
\begin{equation}
a=\frac{\pi}{2} \hskip 10pt \mbox{when}\hskip 10pt \Lambda=|m|\ne0,
\end{equation}
\begin{equation}
a=(|m|+1)\frac{\pi}{2} \hskip 10pt \mbox{when}\hskip 10pt \Lambda =0,
\end{equation}
\end{subequations}
\begin{subequations}
\begin{equation}
\alpha=\sum_{n=0}^N\alpha^{(2n+1)},
\end{equation}
\begin{equation}
\alpha^{(2n+1)}=\mbox{Re}\frac{1}{2}\int_{\Lambda_\alpha}q^{(2n+1)}(\eta)d\eta,
\end{equation}
\end{subequations}
\begin{subequations}
\begin{equation}
\beta=\sum_{n=0}^N\beta^{(2n+1)},
\end{equation}
\begin{equation}
\beta^{(2n+1)}=-\mbox{Re}\frac{1}{2}\int_{\Lambda_\beta}q^{(2n+1)}(\eta)d\eta,
\end{equation}
\end{subequations}
\begin{subequations}
\begin{equation}
\bar K=\sum_{n=0}^N\bar K_{2n},
\end{equation}
\begin{equation}
\bar K_{2n}=\frac{i}{2\pi}\int_ {\Lambda_{K}} q^{(2n+1)}(\eta)d\eta.
\end{equation}
\end{subequations}
For the super-barrier case (Fig. 4) we can instead of (3.15b) use 
the alternative formula
\begin{align}
\bar K_{2n}&=-2\mbox{Im}
\frac{1}{2\pi}\int_{\Lambda_\alpha}q^{(2n+1)}(\eta)d\eta\nonumber\\
&= -2\mbox{Im}
\frac{1}{2\pi}\int_{\Lambda_\beta}q^{(2n+1)}(\eta)d\eta,\tag{3.15$b'$}
\end{align}
which is useful in connection with the transformation to complete
elliptic integrals. With the use of (3.13a,b), (3.14a,b) and (3.15a,$b'$)
each one of the quantities $\alpha,\beta$ and $\bar K$ is then obtained 
as the real or imaginary part of an integral over the contour
$\Lambda_\alpha$ or $\Lambda_\beta$. The contours of integration
$\Lambda_\alpha,\Lambda_\beta$ and $\Lambda_K$ for sub-barrier and
super-barrier energies are shown in Fig. 3 and Fig. 4. The analytic
expressions for the quantity $\tilde \phi$ are given in terms of $\bar
K$ and $\bar K_{2n}$ by (A28) and (A29a,b,c) in the Appendix. The
quantity $\tilde \phi$ is of decisive importance for energies in the
neighbourhood of the top of the barrier. The quantities $\alpha$ and
$\beta$ are positive. The quantity $\bar K_0$ is positive when $\eta_2$
and $\eta_3$ are real, it is equal to zero when $\eta_2$ and $\eta_3$
coincide, and it is negative when $\eta_2$ and $\eta_3$ are complex
conjugate. The quantity $\bar K_2$ may be positive or negative
irrespectively of whether $\eta_2$ and $\eta_3$ are real or complex
conjugate.

Analogously as we changed the original contour $ \Lambda_{\tilde L}$ into
$\Lambda_{\tilde L'}$, when dealing with the quantization condition
(3.5a), we can, when $\Lambda=|m|\ne 0$, change the contours
$\Lambda_\alpha$ and $\Lambda_\beta$ depicted in Fig. 3(a) and Fig.
4(a), so that, instead of letting them enclose only $\eta_1,\eta_2$ and
$\eta_3,\eta_4$, respectively, we make each one of them enclose also a 
pole, $\eta=-1$ or $\eta=+1$, respectively. Calling the new contours 
$\Lambda_{\alpha '}$ and $\Lambda_{\beta '}$ [see Fig. 3(a)
and Fig. 4(a)] and defining 
\begin{subequations}
\begin{equation}
\alpha'=\sum_{n=0}^N\alpha^{'(2n+1)},
\end{equation}
\begin{equation}
\alpha^{'(2n+1)}=\mbox{Re}\frac{1}{2}\int_{\Lambda_{\alpha '}}
q^{(2n+1)}(\eta)d\eta,
\end{equation}
\end{subequations}
\begin{subequations}
\begin{equation}
\beta '=\sum_{n=0}^N\beta ^{'(2n+1)},
\end{equation}
\begin{equation}
\beta ^{'(2n+1)}=-\mbox{Re}\frac{1}{2}\int_{\Lambda_{\beta '}}
q^{(2n+1)}(\eta)d\eta,
\end{equation}
\end{subequations}
and recalling that the functions $Y_{2n}$ are regular analytic at $\eta
= \pm 1$ when $\Lambda =|m| \ne 0$, we find with the use of residue
calculus that for $\Lambda =|m| \ne 0$
\begin{subequations}
\begin{eqnarray}
\alpha ' - \alpha&=&\beta ' -\beta\nonumber\\
&=&\Lambda\pi/2,
\end{eqnarray}
\begin{eqnarray}
\alpha ' -\beta ' & = &\alpha-\beta\nonumber\\
&=&-\frac{b\pi}{2p}.
\end{eqnarray}
\end{subequations}
With the aid of (3.18a,b) we obtain from (3.11) along with (3.12a) for
$\Lambda=|m|\ne 0$ the following quantization condition
\begin{equation}
\mbox{cos}[\alpha '+\beta '+\tilde \phi -(|m|+1)\pi]
={{\mbox{cos}[b\pi/(2p)]}\over {[1+\mbox{exp}(-2\pi\bar K)]^\frac{1}{2}}},
\end{equation}
which has the same merits, relative to the original form of
the quantization condition, i.e., (3.11) with (3.12a), as were
mentioned in connection with the quantization condition (3.6),
pertaining to the $\xi$-equation.
When $\Lambda =0$ we define 
\begin{subequations}
\begin{equation}
\alpha '=\sum_{n=0}^N\alpha ^{'(2n+1)},
\end{equation}
\begin{equation}
\alpha ^{'(2n+1)}=\alpha^{(2n+1)},
\end{equation}
\end{subequations}
\begin{subequations}
\begin{equation}
\beta '=\sum_{n=0}^N\beta^{'(2n+1)},
\end{equation}
\begin{equation}
\beta ^{'(2n+1)}=\beta^{(2n+1)},
\end{equation}
\end{subequations}
and note that (3.18a) is obviously valid also for $\Lambda =0$. With the use
of the theory of complex integration one finds that also (3.18b) is valid
for $\Lambda =0$. Then it follows
from (3.11), (3.12b) and (3.18a,b) that the quantization condition (3.19)
is valid also when $\Lambda =0$.  The quantization condition (3.19) thus
covers in a unified and convenient form both cases $\Lambda =|m|\ne 0$ and 
$\Lambda =0$.

The quantization condition (3.19) can be rewritten as
\begin{eqnarray}
&&\alpha '+\beta '+\tilde \phi -(|m|+1)\pi\nonumber\\
&&\hskip 30pt=\pm
 \mbox{arccos}{{\mbox{cos}[b\pi/(2p)]}\over 
{[1+\mbox{exp}(-2\pi\bar K)]^\frac{1}{2}}}+2s'\pi,
\end{eqnarray}
where $s'$ is an integer. As already mentioned,
(3.18b) is valid for $\Lambda =|m|\ne 0$ as well as for $\Lambda =0$.
When the plus sign in (3.22) applies, we use (3.18b) to express $\beta
'$ in terms of $\alpha '$, and when the minus sign in (3.22) applies, we
use (3.18b) to express $\alpha '$ in terms of $\beta '$. Replacing 
$s'$ by $s_\alpha$ or $s_\beta$, we thus obtain from (3.22) and (3.18b) the two
quantization conditions
\begin{subequations}
\begin{eqnarray}
\alpha '&=&\left(\frac{|m|}{2}+s_\alpha +\frac{1}{2}\right)\pi
-\frac{\tilde \phi}{2} -\frac{b\pi}{4p}\nonumber\\
&&+\frac{1}{2}\mbox{arccos}{{\mbox{cos}\left(\frac{b\pi}{2p}\right)}
\over {[1+\mbox{exp}(-2\pi\bar K)]^\frac{1}{2}}},
\end{eqnarray}
\begin{eqnarray}
\beta '&=&\left(\frac{|m|}{2}+s_\beta +\frac{1}{2}\right)\pi
-\frac{\tilde \phi}{2} +\frac{b\pi}{4p}\nonumber\\
&&-\frac{1}{2}\mbox{arccos}{{\mbox{cos}\left(\frac{b\pi}{2p}\right)}\over 
{[1+\mbox{exp}(-2\pi\bar K)]^\frac{1}{2}}},
\end{eqnarray}
\end{subequations}
where we choose the branch of \mbox{arccos} such that the last two terms on the
right-hand side of (3.23a,b) cancel in the limit $\bar K\rightarrow
+\infty$, that is, when the barrier becomes infinitely thick. 
In this limit, the formulas (3.23a,b) simplify to 
\begin{subequations}
\begin{equation}
\alpha '=\left(\frac{|m|}{2}+s_\alpha +\frac{1}{2}\right)\pi,
\end{equation}
\begin{equation}
\beta '=\left(\frac{|m|}{2}+s_\beta +\frac{1}{2}\right)\pi.
\end{equation}
\end{subequations}

For the particular case that we have a symmetric two-centre Coulomb problem,
i.e., that $Z_1=Z_2$, as is the case for the ion $H_2^+$, the
double-well potential pertaining to the $\eta$-equation becomes
symmetric ($b=0$), and the quantization conditions (3.23a,b) can be simplified:
\begin{subequations}
\begin{eqnarray}
\alpha '&=&\left(\frac{|m|}{2}+s_\alpha +\frac{1}{2}\right)\pi
-\frac{\tilde \phi}{2} \nonumber\\
&&+\frac{1}{2}\mbox{arctan} \mbox{ exp}(-\pi\bar K),
\end{eqnarray}
\begin{eqnarray}
\beta '&=&\left(\frac{|m|}{2}+s_\beta +\frac{1}{2}\right)\pi
-\frac{\tilde \phi}{2} \nonumber\\
&&-\frac{1}{2}\mbox{arctan} \mbox{ exp}(-\pi\bar K).
\end{eqnarray}
\end{subequations}
The reduced separation constant $A'$, obtained from (3.23a,b) in the 
general case and from (3.25a,b) in the symmetric case, is a function of $p^2$
and $C$.

\subsubsection{Comments on the quantization conditions}

In the existing semi-classical treatments, the quantization conditions
derived on assumptions valid for $m\ne 0$ are in general extrapolated to
$m=0$ (corresponding to a particularization of our case $\Lambda =0$)
without any motivation. It is, however, not allowed to obtain a quantization
condition corresponding to $m=0$ from a quantization condition
corresponding to $m\ne 0$ by letting $m$ take continuous values and tend
to zero by a limiting procedure. That the formulas obtained by such an
extrapolation are valid is an {\it a posteriori} conclusion. The correct
justification of the quantization conditions for $\Lambda = 0$ rests on
the use of the connection formula in subsection A.2.b of the Appendix.

For given values of $r_{12},m$ and $E$ one can, as already mentioned,
obtain the possible values of the reduced separation constant $A'$ in the
differential equation (2.8a) with (2.9a) by applying phase-integral
quantization conditions for a single-well potential, while to obtain the
possible values of the reduced separation constant $A'$ in the differential
equation (2.8b) with (2.9b) one has to use quantization conditions either for a
single-well potential or for a double-well potential. The appropriate
quantization condition for the $\xi$-equation determines $A'$ as a
function of $p^2$ and $\tilde C$. The appropriate quantization condition
for the $\eta$-equation determines $A'$ as a function of $p^2$ and $C$.
The eigenvalues of $p^2$, and hence the energy eigenvalues $E$, are
obtained from the requirement that these two expressions for $A'$ must
be equal to each other. One then obtains $A'$ from the quantization
condition for the $\xi$-equation or the $\eta$-equation. 
 The value thus obtained for $A'$
depends obviously on the choice of $C$ and $\tilde C$. One should choose
these parameters as funcitons of $r_{12}$ in such a way that very
accurate values of $p^2$ and $A'$ are obtained already in the first
order of the phase-integral approximation. A practically useful
criterion for this can be formulated as follows. For every value of
$r_{12}$ one determines $C$ and $\tilde C$ such that the first-order
approximation gives the same value as the third-order approximation for
$p^2$(and hence for the energy) as well as for $A'$. In this way
one can make all calculations within the framework of the phase-integral
method.

\acknowledgments
The work of M.L. forms part of a Department of Science and Technology, 
Government of India, research project. Support from the Swedish Natural 
Science Research Council for M. Lakshmanan's visits to Uppsala is 
gratefully acknowledged.
\appendix 
\section{Phase-integral method}
Since the present paper is based on phase-integral formulas that are
scattered in different publications, we collect in this Appendix the
background material that is necessary for reading the paper.

The phase-integral method for solving differential equations of the type

\begin{equation}
\frac{d^2\psi}{dz^2}+R(z)\psi =0
\end{equation}
involves the following items:
\begin{enumerate}
\item
Arbitrary-order phase-integral approximation generated from an 
unspecified base function $Q(z)$, as described in Chapter 1 of \cite{fro10};
see also Dammert and P.O. Fr\"oman \cite{dam}.
\item
The method for solving connection problems developed by Fr\"oman and
Fr\"oman \cite{fro8}, generalized to apply to the phase-integral approximation
referred to in the above item.
\item
Supplementary quantities, expressed analytically in terms of
phase-integrals. An example is the quantity $\tilde \phi$, which is a
new notation for the quantity $-2\sigma$ in Fr\"oman {\it et al.}\cite{fro16},
and which is of decisive importance, when two transition zeros lie close
to each other.

\end{enumerate}
We shall first briefly describe the phase-integral approximation referred to in
item 1. Then we collect connection formulas pertaining to a single
transition point [first-order zero or first-order pole of $Q^2(z)$] and to
a real potential barrier, which can be derived by means of the method
mentioned in item 2 combined with comparision equation technique for
obtaining the 
supplementary quantity $\tilde \phi$ mentioned in item 3 and appearing
in the connection formula for a real barrier. Finally we
present quantization conditions for single-well and double-well
potentials, which can be derived by means of the connection formulas
just mentioned. These quantization conditions are used in our treatment
of the two-centre Coulomb problem.
\subsection{Phase-integral approximation generated from an unspecified
base function}

For a detailed description of the phase-integral approximation generated
from an unspecified base function we refer to chapter 1 in \cite{fro10}.
 A brief description is given below.

In the arbitrary-order phase-integral approximation in question there
appears an unspecified function $Q(z)$ called the {\it base function}.
This function is often chosen to be equal to $R^{\frac{1}{2}}(z)$, but
in many physical problems it is important to use the possibility of
choosing $Q(z)$ differently when there exist certain exceptional points,
for example the origin in connection with the radial Schr\"odinger
equation, and, correspondingly, the poles of $\tilde Q^2(\xi)$ and
$Q^2(\eta)$ at $\xi=1$ and $\eta=\pm 1$ in the two-centre Coulomb
problem. In the present paper we introduce in the base functions a
parameter $\Lambda$, chosen such that either $\Lambda=|m|\ne 0$ or
$\Lambda=0$, and two parameters $C$ and $\tilde C$ to be determined such
that the first- and third-order results coincide, in order that the
first-order approximation be as good as possible.

To be able to write the phase-integral approximation generated from an
unspecified base function in condensed form one introduces the new
independent variable
\begin{equation}
\zeta = \int^z Q(z)dz
\end{equation}
and the function
\begin{equation}
\varepsilon_0=Q^{-\frac{3}{2}}(z)\frac{d^2}{dz^2}Q^{-\frac{1}{2}}(z)+{{R(z)
-Q^2(z)}\over{Q^2(z)}}.
\end{equation}
It can be shown that in a local region of the complex $z$-plane, where
the absolute value of $\varepsilon_0$ is small, the differential equation
(A1) has the approximate solutions
\begin{subequations}
\begin{equation}
\psi=q^{-\frac{1}{2}}(z)\mbox{exp}[\pm iw(z)],
\end{equation}
\begin{equation}
w(z)=\int_{z_0}^zq(z)dz,
\end{equation}
\end{subequations}
where the lower limit of integration $z_0$ is an unspecified constant,
and the function $q(z)$, pertaining to the phase-integral approximation
of the order $2N+1$, is given by
\begin{subequations}
\begin{equation}
q(z)=\sum_{n=0}^Nq^{(2n+1)}(z),
\end{equation}
\begin{equation}
q^{(2n+1)}(z)=Q(z)Y_{2n},
\end{equation}
\end{subequations}
with the first few functions $Y_{2n}$ given by
\begin{subequations}
\begin{equation}
Y_0=1,
\end{equation}
\begin{equation}
Y_2=\frac{1}{2}\varepsilon_0,
\end{equation}
\begin{equation}
Y_4=-\frac{1}{8}\varepsilon_0^2-\frac{1}{8}\frac{d^2\varepsilon_0}{d\zeta^2}.
\end{equation}
\end{subequations}
The choice of the function $Q(z)$ does not affect the expressions for
$Y_{2n}$ in terms of $\varepsilon_0$ and $\zeta$; only the expressions for
$\varepsilon_0$ and $\zeta$ as functions of $z$ are affected.

It is an essential advantage of the phase-integral 
approximation described
above versus the Carlini\cite{car} (JWKB) approximation in higher order that the
former approximation contains the unspecified base function $Q(z)$,
which one can take advantage of in several ways. A criterion for the
determination of the base function is that the function $\varepsilon_0$ be
in some sense small in the region of the complex $z$-plane that is relevant for
the problem under consideration. However, this criterion does not
determine the base function $Q(z)$ uniquely; it turns out that, within
certain limits, the results are not very sensitive to the choice of
$Q(z)$, when the approximation is used in higher orders. With
a convenient choice of $Q(z)$ already the first-order approximation can
be very good. On the other hand an
inconvenient, but possible, choice of $Q(z)$ introduces in the
first-order approximation an unnecessarily large error that, however,
in general becomes corrected already in the third-order approximation. 

The freedom that one has in the choice of the base function $Q(z)$ will
be illuminated in a concrete way below. For a radial Schr\"odinger
equation the usual choice of $Q^2(z)$ is
\begin{subequations}
\begin{equation}
Q^2(z)=R(z)-\frac{1}{4z^2}.
\end{equation}
However, the replacement of (A7a) by
\begin{equation}
Q^2(z)=R(z)-\frac{1}{4z^2}-\frac{\mbox{const}}{z},
\end{equation}
where the coefficient of $\frac{1}{z}$ should be comparatively small, does
not destroy the great accuracy of the results usually obtained with the
phase-integral approximation in higher orders. There is thus a whole set
of base functions that may be used, and there are various ways in which
one can take advantage of this nonuniqueness to make the choice of the
base function well adapted to the particular problem under
consideration. For instance, by adapting the choice of $Q^2(z)$ to the
analytical form of $R(z)$ one can sometimes achieve the result that the
integrals occurring in the phase-integral approximation can be evaluated
analytically. To give an example we assume that $R(z)$ contains only
\mbox{exp}$(z)$ but not $z$ itself. In this case it is convenient to replace
the choice (A7b) by the choice
\begin{equation}
Q^2(z)=R(z)-\frac{1}{4(e^z-1)^2}-\frac{\mbox{const}}{e^z-1}.
\end{equation}
By a convenient choice, for instance of the unspecified
coefficient in (A7b) or (A7c), one can sometimes attain the result
that, for example, eigenvalues or phase-shifts are obtained exactly for
some particular parameter value in every order of the phase-integral
approximation. By making this exactness fulfilled in the limit of a
parameter value, for which the phase-integral result without this
adaptation would not be good, one can actually extend the region of
validity of the phase-integral treatment; see p. 12 in \cite{fro13}.
When the differential equation
contains one or more parameters, the accurate calculation of the wave
function may require different choices of the base function $Q(z)$ for
different ranges of the parameter values. To illustrate this fact we
consider the radial Schr\"odinger equation. For sufficiently large
values of the angular momentum quantum number $l$ we obtain an accurate
phase-integral approximation (valid also close to $z=0$) if we choose
$Q^2(z)$ according to (A7a) or (A7b). If the value of $l$ is too
small, this phase-integral approximation is not good. It can be
considerably improved (except close to $z=0$), when the absolute value of
the coefficient of $\frac{1}{z}$ in $R(z)$ is sufficiently large, if one
chooses instead
\begin{equation}
Q^2(z)=R(z)+\frac{l(l+1)}{z^2}.
\end{equation}
\end{subequations}
The corresponding phase-integral approximation is not valid close to
$z=0$, but the wave function that is regular and tends to $z^{l+1}$ as
$z\rightarrow 0$ can be obtained sufficiently far away from $z=0$ by
means of the connection formula that will be presented in subsection A.2.b
of this Appendix. 

The appearance of the unspecified base function $Q(z)$ in the
phase-integral approximation is thus very important from several points
of view. In our treatment of the two-centre Coulomb problem we use two
essentially different kinds of base function [corresponding to 
 $\Lambda =|m|\ne 0$ and $\Lambda
=0$ in (3.2a,b)], which yield approximate
solutions with different regions of validity.

When the first-order approximation is used, it is often convenient to
choose the constant lower limit of integration $z_0$ in (A4b) to be a
zero or a first-order pole of $Q^2(z)$. This is, however, in general
not possible when a higher-order approximation is used, since the
integral in (A4b) would then in general 
be divergent. If $z_0$ is an odd-order zero or an odd-order pole of
$Q^2(z)$, it is therefore convenient to replace the definition (A4b) of
$w(z)$ by the definition

\begin{equation}
w(z)=\frac{1}{2}\int_{\Gamma_{z_0}(z)}q(z)dz,
\end{equation}
where $\Gamma_{z_0}(z)$ is a path of integration that starts at the
point corresponding to $z$ on a Riemann sheet adjacent to the complex
$z$-plane under consideration, encircles $z_0$ in the positive or in the
negative sense, and ends at $z$. It is immaterial for the value of the
integral in (A8) if the path of integration encircles $z_0$ in the
positive or in the negative sense, but the endpoint must be $z$. 
For the first-order approximation the definitions (A4b) and (A8) are
identical.

It is useful to introduce a short-hand notation for the integral in the
right-hand member of (A8) by the definition
\begin{equation}
\int_{(z_0)}^zq(z)dz=\frac{1}{2}\int_{\Gamma_{z_0}(z)}q(z)dz.
\end{equation}
For the first order of the phase-integral approximation one can replace
$(z_0)$ by $z_0$ in the left-hand member of (A9) and thus get an
ordinary integral from $z_0$ to $z$ instead of half of the integral
along the contour $\Gamma_{z_0}(z)$. In analogy to (A9) one defines a
short-hand notation for an integral in which the upper limit of
integration is an odd-order zero or an odd-order pole of $Q^2(z)$. When
one has two transition points of that kind as limits of integration,
 the definition of the
short-hand notation with both limits within parentheses implies that the
integral is equal to half of the integral along a closed loop enclosing
both transition points. The simplified notation in the left-hand member
of (A9) for the integral in the right-hand member of (A9) was
introduced by Fr\"oman {\it et al.} \cite{fro14}, pp 160-161. It makes it
possible to use, for an arbitrary order of the phase-integral
approximation, a similar simple notation and almost the same simple
language (although in a generalized sense) as for the first order of the
phase-integral approximation. One thus achieves a great formal and
practical simplification in the treatment of concrete problems, when an
arbitrary order of the phase-integral approximation is used.

We remark that the notations used above differ from the notations in the
original papers published up to the beginning of the eighties in the
respect that $Q^2(z)$ and $Q_{mod}^2(z)$ in those papers correspond in
the present paper to $R(z)$ and $Q^2(z)$, respectively.

\subsection{Connection formulas associated with a single transition
point}
\subsubsection{Connection formulas pertaining to a first-order, real
transition zero}
Before the phase-integral approximation generated from an unspecified
base function had been introduced, N. Fr\"oman \cite{fro5} derived
arbitrary-order connection formulas associated with a turning
point for the particular phase-integral approximation
of arbitrary order corresponding to $Q^2(z)=R(z)$. After the
phase-integral approximation generated from an unspecified base function
had been introduced, it turned out that these connection formulas remain
valid also when $Q^2(z)\ne R(z)$. Below we shall present the general
connection formulas.

The functions $R(z)$ and $Q^2(z)$ are assumed to be real on the real
$z$-axis (the $x$-axis). On this axis there is a generalized classical
turning point $t$, i.e., a simple zero of $Q^2(z)$, and there is, in a
generalized sense, a classically allowed region on one side of $t$,
i.e., a region where $Q^2(x)>0$, and a classically forbidden region on
the other side of $t$, i.e., a region where $Q^2(x)<0$. Defining
\begin{equation}
w(x)=\int_{(t)}^xq(z)dz,
\end{equation}
we can write the connection formula for tracing a phase-integral
solution of (A1) from the classically allowed to the classically
forbidden region as
\begin{eqnarray}
&&A\left |q^{-\frac{1}{2}}(x)\right |\mbox{exp}\left \{i\left [\left
|w(x)\right |+\frac{\pi}{4}\right]\right\}\nonumber\\
&&\hskip 20pt+B\left |q^{-\frac{1}{2}}(x)
\right |\mbox{exp}\left \{-i\left [\left
|w(x)\right |+\frac{\pi}{4}\right]\right\}\longrightarrow\nonumber\\
&&\hskip 40pt\longrightarrow
(A+B)\left |q^{-\frac{1}{2}}(x)\right |\mbox{exp} \left [
|w(x) |\right],
\end{eqnarray}
where $A$ and $B$ are constants, which are arbitrary except for the
requirement that $\frac{A+B}{|A|+|B|}$ must not be too close to
zero. As a consequence of (A11) we have the connection formula
\begin{eqnarray}
&&\left |q^{-\frac{1}{2}}(x)\right |\mbox{cos}\left [\left
|w(x)\right |+\delta-\frac{\pi}{4}\right]\rightarrow\nonumber\\
&&\hskip 20pt \rightarrow \mbox{sin} \mbox{ }\delta
\left |q^{-\frac{1}{2}}(x)\right |\mbox{exp}\left [\left
|w(x)\right |\right],
\end{eqnarray}
where $\delta$ is a real phase constant that must not be too close to
zero. The connection formula for tracing a phase-integral solution of
(A1) from the classically forbidden to the classically allowed region is
\begin{eqnarray}
&&\left |q^{-\frac{1}{2}}(x)\right |\mbox{exp}\left [ -
|w(x)|\right]+C\left |q^{-\frac{1}{2}}(x)\right |\mbox{exp}\left [\left
|w(x)\right |\right]\rightarrow \nonumber\\
&&\hskip 20pt \rightarrow 2\left |q^{-\frac{1}{2}}(x)\right
|\mbox{cos}\left [\left|w(x)\right|-\frac{\pi}{4}\right],
\end{eqnarray}
and it is valid provided that the condition
\begin{equation}
C\mbox{exp}\left[\left|w(x)\right|\right] \lesssim\mbox{exp}
\left[-\left|w(x)\right|\right],
\end{equation}
is fulfilled. For a numerical study of the accuracy and the properties 
of the connection formula (A13) with $C=0$ we refer to N. Fr\"oman and
Mrazek\cite{fro11}. We emphasize the one-directional character of the
connection formulas (A11), (A12) and (A13), which means that the tracing
of a solution must always be made in the direction of the arrow. This
property of the connection formulas has been thoroughly investigated and
even illustrated numerically by N. Fr\"{o}man \cite{fro4a} for the first order
of the Carlini \cite{car} (JWKB) approximation. The whole discussion in 
\cite{fro4a}
applies in principle to the connection formulas for the higher orders of
the phase-integral approximation as well. The above connection formulas
for the phase-integral approximation of any order may in many cases be
used for obtaining very accurate solutions of physical problems, when
the classical turning points are well separated, and when there are no
other transition points near the real axis in the region of the complex
$z$-plane of interest. Within their range of applicability the
connection formulas are very useful because of their simplicity and the
great ease with which they can be used.

\subsubsection{Connection formula pertaining to a first-order transition
pole}
Now we assume that in a certain region of the complex $z$-plane around a
first-order transition pole $t$, i.e., a first-order pole of $Q^2(z)$,
we have
\begin{equation}
R(z)=-\frac{l(l+1)}{(z-t)^2}+\frac{B}{z-t}+\mbox{ a regular function of
}z,
\end{equation}
\begin{equation}
Q^2(z)=\frac{\bar B}{z-t}+\mbox{ a regular function of }z,
\end{equation}
where $2l+1$ is a non-negative integer. We assume that the absolute
values of $B$ and $\bar B$ are sufficiently large, while the absolute value
of $B-\bar B$ and the absolute value of the difference between the two regular functions in (A15)
and (A16) are sufficiently small. There is one particular curve on which $w(z)$,
defined as
\begin{equation}
w(z)=\int_{(t)}^zq(z)dz,
\end{equation}
is real. For the first order of the phase-integral approximation this is
the anti-Stokes line that emerges from $t$. Therefore we use, also when
a higher order of the phase-integral approximation is used, the
terminology an {\it  anti-Stokes line that emerges from $t$} in a
generalized sense to denote the anti-Stokes line on which $w(z)$, defined in
(A17), is real. For the first-order approximation (and under certain
unnecessarily restrictive assumptions) Fr\"oman and Fr\"oman \cite{fro8}
obtained a phase-integral formula [their eq.(7.28)], valid sufficiently
far away from $t$ on the anti-Stokes line that emerges from $t$, for the
particular solution $\psi (z)$ of (A1) that fulfils the condition
\begin{equation}
\lim_{z\rightarrow t}\frac{\psi(z)}{(z-t)^{l+1}}=1.
\end{equation}
This formula can be generalized to be valid for an arbitrary order of
the phase-integral approximation generated from an unspecified base
function and can then be formulated as follows. On the lip of the
anti-Stokes line emerging from $t$, where {\it arg} $w(z)$ is smallest, 
the solution of (A1) that fulfils the condition (A18) is, sufficiently far 
away from $t$, given by the phase-integral formula
\begin{equation}
\psi(z)=\left (\pi
c\frac{w(z)}{|w(z)|}\right )^{-\frac{1}{2}}q^{-\frac{1}{2}}(z)
\mbox{cos}\left[|w(z)|-\left(l+\frac{3}{4}\right)\pi\right],
\end{equation}
where $c$ is the residue of [$\psi (z)]^{-2}$ at $z=t$ and is thus
determined by the expansion of $\psi (z)$ in powers of $z-t$, and
the sign of $\left [\pi c\frac{w(z)}{|w(z)|}\right ]^{-\frac{1}{2}}$ 
has to be chosen conveniently. For the
special case that $l=0$ one finds that $c=B$. Formula (A19) can easily
be particularized to the case that $R(z)$ and $Q^2(z)$ are real on the
real $z$-axis (the $x$-axis) and $t$ lies on that axis. 

\subsection{Connection formula for a real, smooth, single-hump potential
barrier}
Our starting point is a paper by Fr\"oman and Fr\"oman \cite{fro9}. Although
it was assumed in the treatment there that $Q^2(z)=R(z)$ with the
notations in the present paper, the results
obtained are valid also when $Q^2(z)\ne R(z)$. In the present paper it
is convenient to introduce partly other notations than in 
\cite{fro9}. Thus we now denote by $t'$ and $t''$ the two relevant 
zeros of $Q^2(z)$, i.e., the two generalized classical turning points in
the sub-barrier case $(t'<t'')$ and the two complex conjugate transition zeros
in the super-barrier case (Im $t'\le 0,\mbox{Im }t''\ge 0$). Let
$x'$, called $x_1$ in \cite{fro9}, be a
point in the classically allowed region of the real $z$-axis to the left
of the barrier, and let $x''$, called $x_2$ in 
\cite{fro9}, be a point in the classically allowed
region of the real $z$-axis to the right of the barrier. We introduce
the notations
\begin{subequations}
\begin{equation}
\theta =|F_{22}|\mbox{ exp}(K),
\end{equation}
\begin{equation}
\vartheta = \mbox{arg } F_{22},
\end{equation}
\begin{equation}
\tilde \phi = -2\sigma = \frac{\pi}{2}-\mbox{arg } F_{12},
\end{equation}
\end{subequations}
where $F_{12}$ and $F_{22}$ are defined in 
\cite{fro9}, and $K$ is defined by eq.(12) in 
\cite{fro9}. In the definitions (A20a,b,c) it is
assumed that the phase of $q^\frac{1}{2}(z)$ is chosen as shown in Fig.1 in
\cite{fro9}. We shall, however, in the following write
the formulas in such a way that they remain unchanged if one changes the
phase of $Q^\frac{1}{2}(z)$ and hence the phase of $q^\frac{1}{2}(z)$; see
(A5a,b). The quantity $K$ in (A20a) is then given by
\begin{eqnarray}
K&=&\frac{1}{2}i\int_\Lambda q(z)dz\nonumber\\
&=&i\int_{(t')}^{(t'')}q(z)dz,
\end{eqnarray}
where $\Lambda$ [not to be confused with the parameter $\Lambda$ in the
base function $Q(\eta)$] is a closed contour of
integration encircling both $t'$ and $t''$, but no other transition
point, and the integration is performed in the direction that in the
{\it first-order approximation} yields $K>0$ for energies below the top
of the barrier and $K<0$ for energies above the top of the barrier. If
higher-order approximations are used, the quantity $K$ may become
negative also for energies below (but not too far from) the top of the
barrier; see Table I in \cite{fro7}. We have replaced $\sigma$,
defined by eq. (28) in \cite{fro9}, by $-\frac{\tilde
\phi}{2}$ [cf. (A20c)] in order to get better agreement with a notation
used by other authors; see for instance Child \cite{child}. 
Now we define
\begin{subequations}
\begin{equation}
B'=A_1\mbox{exp}\left(-i\frac{\pi}{4}\right),
\end{equation}
\begin{equation}
A'=B_1\mbox{exp}\left(+i\frac{\pi}{4}\right),
\end{equation}
\end{subequations}
where the notations in the right-hand members are those used in 
\cite{fro9}. Using the short-hand notation 
defined in (A9), we
obtain from eqs. (25a) and (9a), with $x_1$ replaced by $x'$, in 
\cite{fro9} and (A22a,b) in the present paper
\begin{subequations}
\begin{eqnarray}
\psi
(x')&=&A'\left|q^{-\frac{1}{2}}(x')\right|\mbox{exp}\left(+i\left|\mbox{Re}
\int_{(t')}^{x'}q(z)dz\right|\right)\nonumber\\
&&+B'\left|q^{-\frac{1}{2}}(x')\right|\mbox{exp}\left(-i\left|\mbox{Re}
\int_{(t')}^{x'}q(z)dz\right|\right),
\end{eqnarray}
and from eqs. (25b) and (23), with $x_2$ replaced by $x''$, in 
\cite{fro9} and (A20a,b,c) and (A22a,b) in the
present paper
\begin{eqnarray}
\psi
(x'')&=&A''\left|q^{-\frac{1}{2}}(x'')\right|\mbox{exp}\left(+i\left|\mbox{Re}
\int_{(t'')}^{x''}q(z)dz\right|\right)\nonumber\\
&&+B''\left|q^{-\frac{1}{2}}(x'')
\right|\mbox{exp}\left(-i\left|\mbox{Re}
\int_{(t'')}^{x''}q(z)dz\right|\right),
\end{eqnarray}
\end{subequations}
where
\begin{equation}
\begin{pmatrix}
A''\\
B''
\end{pmatrix}
= \tilde M
\begin{pmatrix}
A'\\
B'
\end{pmatrix},
\end{equation}
\begin{subequations}
\begin{equation}
\tilde M =
\begin{pmatrix}
\theta\mbox{exp}\left [-i(\frac{\pi}{2}+\vartheta)\right] &
(\theta^2+1)^{\frac{1}{2}}\mbox{exp}(+i\tilde\phi)\\
(\theta^2+1)^{\frac{1}{2}}\mbox{exp}(-i\tilde\phi) &
\theta\mbox{exp}\left [+i(\frac{\pi}{2}+\vartheta)\right]
\end{pmatrix},
\end{equation}
\begin{equation}
\mbox{det}\tilde M =-1.
\end{equation}
\end{subequations}
It is seen from (A23a,b) that the coefficients $A'$ and $A''$ are
associated with waves travelling away from the barrier, while the
coefficients $B'$ and $B''$ are associated with waves incoming towards
the barrier. From (A25a,b) we obtain
\begin{equation}
\tilde M^{-1} =
\begin{pmatrix}
\theta\mbox{exp}\left [-i(\frac{\pi}{2}-\vartheta)\right] &
(\theta^2+1)^{\frac{1}{2}}\mbox{exp}(+i\tilde\phi)\cr
(\theta^2+1)^{\frac{1}{2}}\mbox{exp}(-i\tilde\phi) &
\theta\mbox{exp}\left [+i(\frac{\pi}{2}-\vartheta)\right]
\end{pmatrix}.
\end{equation}
One thus obtains $\tilde M^{-1}$ from $\tilde M$ by replacing
$\vartheta$ by $-\vartheta$. We emphasize that the above formulas are in
principle exact, provided that one knows the quantities $\theta,
\vartheta$ and $\tilde\phi$, which depend slightly on $x'$ and $x''$.
Furthermore, the two transition zeros associated with 
the potential barrier need not lie far away from other possibly existing
transition points; see 
\cite{fro9}. However, when one introduces for $\theta,
\vartheta$ and $\tilde\phi$ the approximate values that will be given
below, the barrier is assumed to be well separated (in the sense just
described) from all transition
points that are not associated with the barrier.

When $A'$ and $B'$ are given constants, associated with a wave function
that is given at the point $x'$, the coefficients $A''$ and $B''$, which
are obtained from (A24) along with (A25a), depend slightly on $x'$ and $x''$
via the quantities $\theta, \vartheta$ and $\tilde\phi$. One obtains the
derivatives of $\psi(x')$ and $\psi(x'')$ from (A23a,b) by considering
$A',B',A''$ and $B''$ $formally$ as constants.

When the transition points that are $not$ associated with the barrier
lie sufficiently far away from $t'$ and $t''$, it follows from eq.(43a)
in \cite{fro9} and (A20a,b) that
\begin{subequations}
\begin{equation}
\theta \approx \mbox{exp}(K),
\end{equation}
\begin{equation}
\vartheta \approx 0.
\end{equation}
\end{subequations}

The quantity $\tilde \phi$ is particularly important when the energy is
close to the top of the barrier, but it is important also for energies
well below the top, if one with the use of higher orders of the
phase-integral approximation wants to obtain very accurate results. In practice
one cannot obtain useful expressions for $\tilde\phi$ from the exact
formula (A20c). Under the assumption that $\frac{d^2R(z)}{dz^2}$ is not
too close to zero at the top of the barrier, Fr\"oman {\it et al.} \cite{fro15}
derived by means of comparision equation technique, adapted to yield
formulas for supplementary quantities in the phase-integral method, an
approximate, but very accurate, formula in the $(2N+1)$th order of the
phase-integral approximation [their eqs. (5.5.30), (5.5.25a-g), (5.4.23)
and (5.4.21)], from which we can obtain the formula
\begin{equation}
\tilde \phi = \mbox{arg}\Gamma\left(\frac{1}{2}+i\bar K\right)-\bar
K\mbox{ ln}|\bar K_0|+\sum_{n=0}^N\phi^{(2n+1)},
\end{equation}
where
\begin{subequations}
\begin{equation}
\phi^{(1)}=\bar K_0,
\end{equation}
\begin{equation}
\phi^{(3)}=-\frac{1}{24\bar K_0},
\end{equation}
\begin{equation}
\phi^{(5)}=-\frac{7}{2880\bar K_0^3}+\frac{\bar K_2}{24\bar K_0^2}
-\frac{\bar K_2^2}{2\bar K_0},
\end{equation}
\end{subequations}
with
\begin{subequations}
\begin{equation}
\bar K_{2n}=\frac{1}{2\pi i}\int_{\Lambda}Y_{2n}Q(z)dz, \hskip 10pt
n=0,1,2,....N,
\end{equation}
\begin{equation}
\bar K= \sum_{n=0}^N\bar K_{2n}=\frac{K}{\pi},
\end{equation}
\end{subequations}
$\Lambda$ being the previously described contour of integration
encircling $t'$ and $t''$ but no other transition point, with the
integration performed in the direction that makes $\bar K_0$ positive
when $t'$ and $t''$ are real, i.e., when the barrier is superdense, but
negative when $t'$ and $t''$ are complex conjugate, i.e., when the barrier is
underdense. 
(Note that we perform the integrations in (A21) and (A30a) in opposite
directions in order to make these formulas to agree with eq. (12) in
\cite{fro9} and eq. (5.4.21) in \cite{fro15}, respectively.)
The
result given by (A28), (A29a,b,c) and (A30a,b) can also be obtained from
\cite{fro16}, where $-2\sigma$ is the same 
as our $\tilde \phi$.

We emphasize again that for the validity of (A28) with the expressions (A29a-c)
for $\phi ^{(2n+1)}$ the essential restriction is that
$\left|\frac{d^2R(z)}{dz^2}\right|$ must not be too small at the top of
the barrier, which means that close to its top the barrier is
approximately parabolic. However, when the energy is close to the top of the
barrier, it is the slight deviation from parabolic shape close to the top of
the barrier that determines the values of the quantities $\bar K_{2n},
n>0$, and one needs accurate values of these quantities for obtaining
accurate values of $\tilde \phi$ in higher orders of the phase-integral
approximation.

The barrier connection formula presented in this section is valid
uniformly for all energies, below and above the top of the barrier. We
would also like to emphasize that while the connection formulas
pertaining to an isolated turning point (N. Fr\"oman \cite{fro5}) are
one-directional, the barrier connection formula (A23a,b) with
(A24) and (A25a), which is valid when the barrier is well isolated and
has an approximately parabolic top, is bi-directional. However, the
neighbourhood of an energy that corresponds to a resonance requires a
careful discussion.
\subsection{Quantization conditions for single-well and double-well
potentials}
In this section we shall present quantization conditions for general
single-well potentials \cite{fro4c,fro6,paul} and double-well
potentials \cite{fro16,paul,fro4b,fro12},
 valid for any conveniently chosen order of the
phase-integral approximation, in forms especially adapted to the
two-centre Coulomb problem. In the quantization conditions pertaining to
the double-well potential there appears the supplementary quantity
$\tilde \phi$, which was discussed in Section A.3, and which is of
particular importance for energy eigenvalues in the neighbourhood of the
top of the barrier; cf. numerical results in \cite{fro16},
 where $Q^2,Q_{mod}^2$ and $\sigma$ correspond to $R, Q^2$
and $-\frac{\tilde \phi}{2}$, respectively, in the present paper.
Comparision with numerical results \cite{fro16} shows
that all energy eigenvalues, also the low-lying ones and those in the
neighbourhood of the top of the barrier, are obtained very accurately
from the phase-integral quantization conditions when the third- or
fifth-order approximation is used.

Since arbitrary-order phase-integral quantization conditions for the
single-well and for the double-well potential problem have been given in
previous work, we restrict ourselves to quoting those results and taking
into account the fact that we are dealing with the special potentials
pertaining to the two-centre Coulomb problem.

\subsubsection{Quantization conditions for single-well potentials}
We assume that $R(z)$ and $Q^2(z)$ are real on the real $z$-axis (the
$x$-axis) and that there are two transition points $t'$ and $t''$ $(>t')$
on this axis, each one of which may be either a first-order transition
zero, i.e., a first-order zero of $Q^2(z)$, or a first-order transition
pole, i.e., a first-order pole of $Q^2(z)$. These transition points
are assumed to lie far away from all other transition points. 
 On the real axis between
$t'$ and $t''$ it is assumed that $Q^2(x)$ is positive. With the aid of
the connection formulas in Section A.2 we can derive the quantization
conditions that will be presented below.

When both $t'$ and $t''$ are first-order transition zeros, we obtain the
quantization condition \cite{fro4c,fro6} 
\begin{equation}
\left|\int_{(t')}^{(t'')}q(z)dz\right|=\left (s+\frac{1}{2}\right) \pi, \hskip
10pt s=0,1,2...
\end{equation}

When one of the transition points $t'$ and $t''$ is a first-order
transition zero, and the other is a first-order transition pole in the
neighbourhood of which $R(z)$ and $Q^2(z)$ can be expanded according to
(A15) and (A16) with $l=\frac{|m|-1}{2}$, i.e.,
$l(l+1)=\frac{m^2-1}{4}$, we obtain under the assumptions introduced below
those expansions the quantization condition \cite{fro6}
\begin{equation}
\left|\int_{(t')}^{(t'')}q(z)dz\right |=\left (\frac{|m|}{2}+s+\frac{1}{2}\right) \pi, 
\end{equation}
where $m$ is an integer (positive, negative or zero), and $s$ is also an
integer.

When both transition points $t'$ and $t''$ are first-order transition
poles, in the neighbourhood of which $R(z)$ and $Q^2(z)$ can be expanded
according to (A15) and (A16) with $l=\frac{|m|-1}{2}$, i.e.,
$l(l+1)=\frac{m^2-1}{4}$ for both transition poles but possibly with
different coefficients $B$ and $\bar B$ for the two transition poles, we obtain
under the assumptions introduced below (A15) and (A16) the
quantization condition \cite{fro13}
\begin{equation}
\left|\int_{(t')}^{(t'')}q(z)dz\right|=\left (|m|+s+\frac{1}{2}\right) \pi, 
\end{equation}
where $m$ is an integer (positive, negative or zero), and $s$ is also an
integer.

\subsubsection{Quantization conditions for double-well potentials}
We assume that $R(z)$ and $Q^2(z)$ are real on the real $z$-axis (the
$x$-axis) and that there are either two generalized classical turning
points $t'$ and $t''\mbox{ }(>t')$ [real, simple zeros of $Q^2(z)$] 
associated with a superdense potential barrier or two complex conjugate 
simple transition zeros $t'$ and $t''$ [simple zeros of $Q^2(z)$; Im $t'<0$,
Im $t''>0$] associated with an underdense potential barrier. The
classically allowed region to the left of the
barrier is to the left delimited by a transition point $t_{-}$ (on the
real axis), and the classically allowed region to
the right of the barrier is to the right delimited by a transition point
$t_{+}$ (on the real axis), where  $t_{-}$ and $t_{+}$ are both either
generalized classical turning points, i.e., simple zeros of $Q^2(z)$, 
or first-order transition poles, i.e., first-order poles of $Q^2(z)$.
The points $t_-$ and $t_+$, as well as other possibly existing transition
points that are not associated with the barrier, are assumed to lie far away
from $t'$ and $t''$. When $t_{-}$ and $t_{+}$ are simple
transition poles we assume that in the neighbourhood of $t_{\pm}$ we have
\begin{equation}
R(z)=\frac{(1-m^2)}{4(z-t_{\pm})^2}+\frac{B_{\pm}}{z-t_{\pm}}+\mbox{
a regular function of } z,
\end{equation}
\begin{equation}
Q^2(z)=\frac{\bar B_{\pm}}{z-t_{\pm}}+\mbox{a regular function of }z,
\end{equation}
where $m$ is an integer (positive, negative or zero), $B_{\pm}$ and
$\bar B_{\pm}$ are sufficiently large to their absolute values, while the
absolute values of $B_{\pm}-\bar B_{\pm}$, as well as the difference between the 
regular functions in (A34) and (A35), is not too large. Under the above
assumptions we obtain with the aid of connection formulas in Section
A.2 and Section A.3 the quantization condition 
\begin{eqnarray}
\label{a36}
&&\mbox{tan}\left(\left|\mbox{Re}\int_{(t_{-})}^{(t')}q(z)dz\right|+\frac{\tilde
\phi}{2}-a\right)\mbox{tan}\left(\left|\mbox{Re}\int_{(t_{+})}^{(t'')}q(z)
dz\right|\right.\nonumber\\
&&\hskip 40pt\left.+\frac{\tilde
\phi}{2}-a\right)=
\frac{[1+\mbox{exp}(-2K)]^{\frac{1}{2}}-1}{[1+\mbox{exp}(-2K)]^{\frac{
1}{2}}+1},
\end{eqnarray}
where
\begin{equation}
K=\frac{1}{2i}\int_{\Lambda}q(z)dz,
\end{equation}
$\Lambda$ being a closed contour, enclosing $t'$ and $t''$, along
which the integration is performed in the direction that makes the first-order
contribution to $K$ positive when the barrier is super-dense but negative when
the barrier is under-dense, and
\begin{equation}
a=
\left\{\begin{matrix}
\frac{\pi}{2} & \mbox{when } t_- \mbox{ and } t_+ \mbox{ are transition
zeros},\cr
(|m|+1)\frac{\pi}{2} & \mbox{when } t_- \mbox{ and } t_+ \mbox{ are transition
poles.}
\end{matrix}
\right.
\end{equation}
The quantization condition (\ref{a36}) can be rewritten into the form 
\begin{eqnarray}
&&\mbox{cos}\left(\left|\mbox{Re}\int_{(t_{-})}^{(t')}q(z)dz\right|
+\left|\mbox{Re}\int_{(t_{+})}^{(t'')}q(z)dz\right|+\tilde
\phi-2a\right)\nonumber\\
&&\hskip 10pt ={{\mbox{cos}\left(\left|\mbox{Re}\int_{(t_{-})}^{(t')}q(z)dz\right|
-\left|\mbox{Re}\int_{(t_{+})}^{(t'')}q(z)dz\right|\right)}\over
{[1+\mbox{exp}(-2K)]^\frac{1}{2}}}. 
\end{eqnarray}

\begin{figure}[!ht]
\centerline{\epsfig{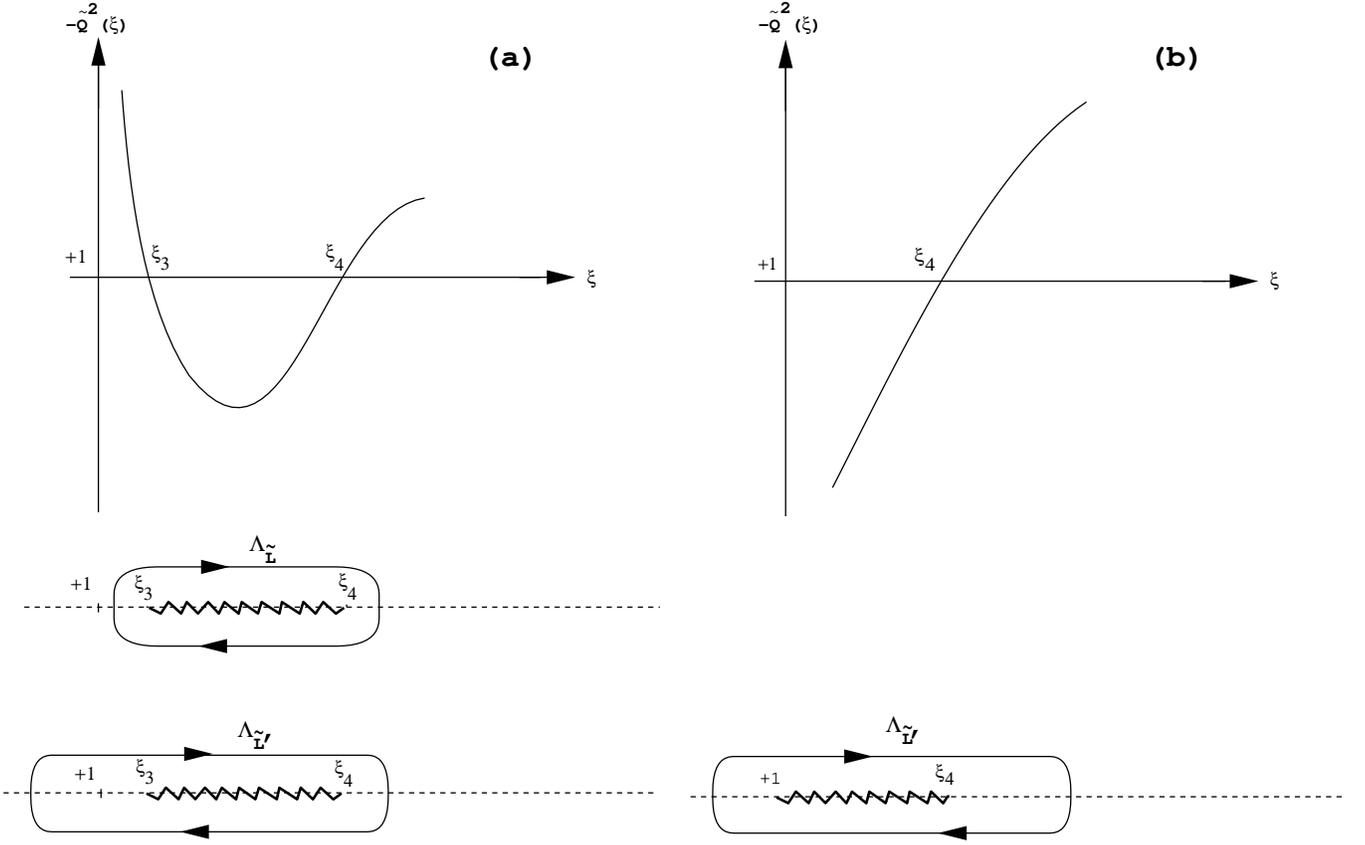}}
\caption
{ 
The figure gives schematic pictures of $-\tilde Q^2(\xi)$ for $\xi >1$ and 
of the contours of integration in the complex $\xi$-plane. The
cuts are indicated by wavy lines. On the upper lip of the cut
 $\tilde Q(\xi)$ is positive. Only those zeros of $\tilde
Q^2(\xi)$ that are relevant for the contours of integration are
shown. For $\Lambda =|m| \ne 0$ (a) always applies, and the relation
between the integrals associated with the contours $\Lambda_{\tilde L}$ and
$\Lambda_{\tilde L'}$ is $\tilde L'=\tilde L+\frac{|m|}{2}$; the zeros
$\xi_1$ and $\xi_2$ of $\tilde Q^2(\xi)$, which are not shown in the figure,
may be real or complex conjugate. For $\Lambda=0$ there are only two
zeros, $\xi_3$ and $\xi_4$, of $\tilde Q^2(\xi)$, and either (a) or (b) 
may apply.}
\label{Fig.1(a)}
\end{figure}
\begin{figure}
\centerline{\epsfig{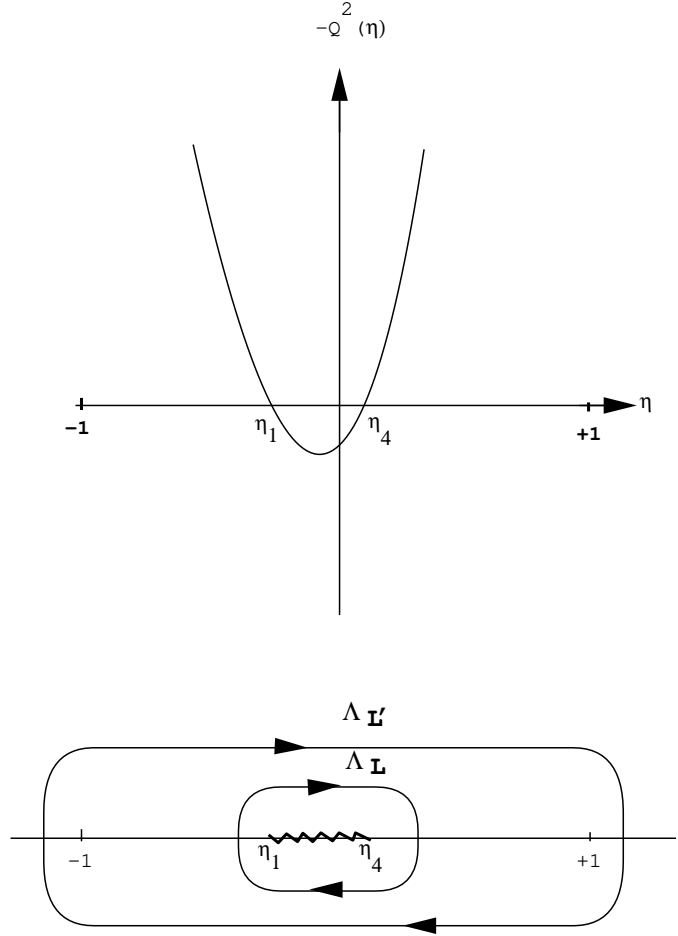}}
\caption
{
The figure gives schematic pictures of $-Q^2(\eta)$
for $-1<\eta<1$ and of the contours of integration when $-Q^2(\eta )$
in the interval $-1<\eta<1$ is a single-well potential, which may occur
for $\Lambda =|m|\ne 0$ as well as for $\Lambda =0$.
The cut is indicated by a wavy line, on
the upper lip of which $Q(\eta)$ is positive. The contour $\Lambda_{L'}$
can be used only when $\Lambda =|m| \ne 0$, and the relation between the integrals
associated with the contours $\Lambda_L$ and 
$\Lambda_{L'}$ is then $L'=L+|m|$. The quantization conditions, expressed
in terms of complete elliptic integrals, for the situation in this figure
with $\Lambda=|m|\ne 0$ are the same as the corresponding quantization 
conditions for the situation in Fig. 4(a). The quantization condition for
the case in the present figure with $\Lambda =0$ has so far not appeared in
the applications.
}
\label{Fig.2(a)}
\end{figure}

\begin{figure}
\centerline{\epsfig{figure=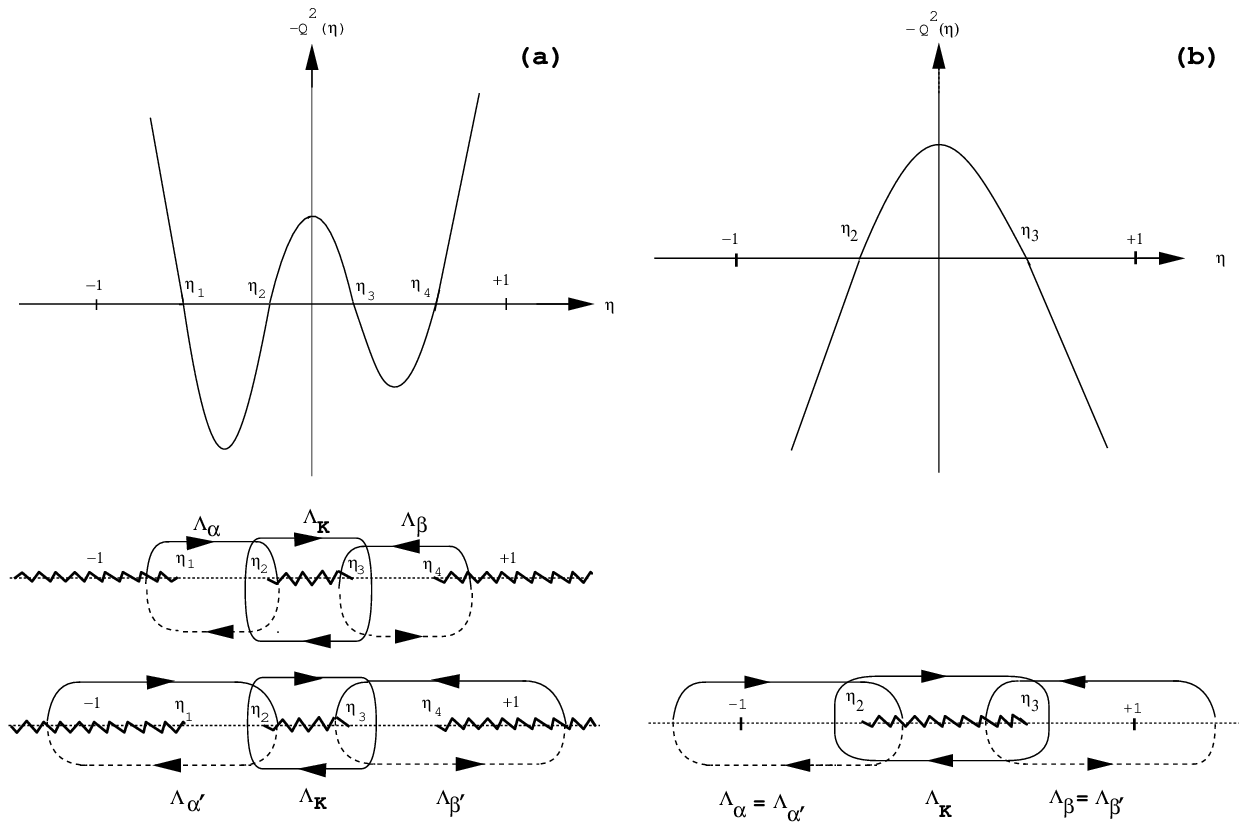, width=\linewidth}}
\caption
{
The figure gives schematic pictures of $-Q^2(\eta)$
for $-1<\eta<1$ and of the contours of integration in the complex
$\eta$-plane, when $-Q^2(\eta)$ corresponds to a double-well potential
with a super-dense barrier. In (a) $\Lambda =|m|\ne 0$ and in (b) $\Lambda =0$.
The cuts are indicated by wavy lines, 
and the parts of the  contours of integration that lie
on  Riemann sheets adjacent to the complex $\eta$-plane under
consideration are dashed. In the left-hand classically allowed region
$Q(\eta)$ is positive. Only those zeros of $Q^2(\eta)$ that are relevant
for the contours of integration are shown. We recall the relations (3.18a) 
which mean that $\alpha'=\alpha+\frac{\Lambda\pi}{2}$ and $\beta'=\beta+
\frac{\Lambda\pi}{2}$.}
\label{Fig.3(a)}
\end{figure}

\begin{figure}
\centerline{\epsfig{figure=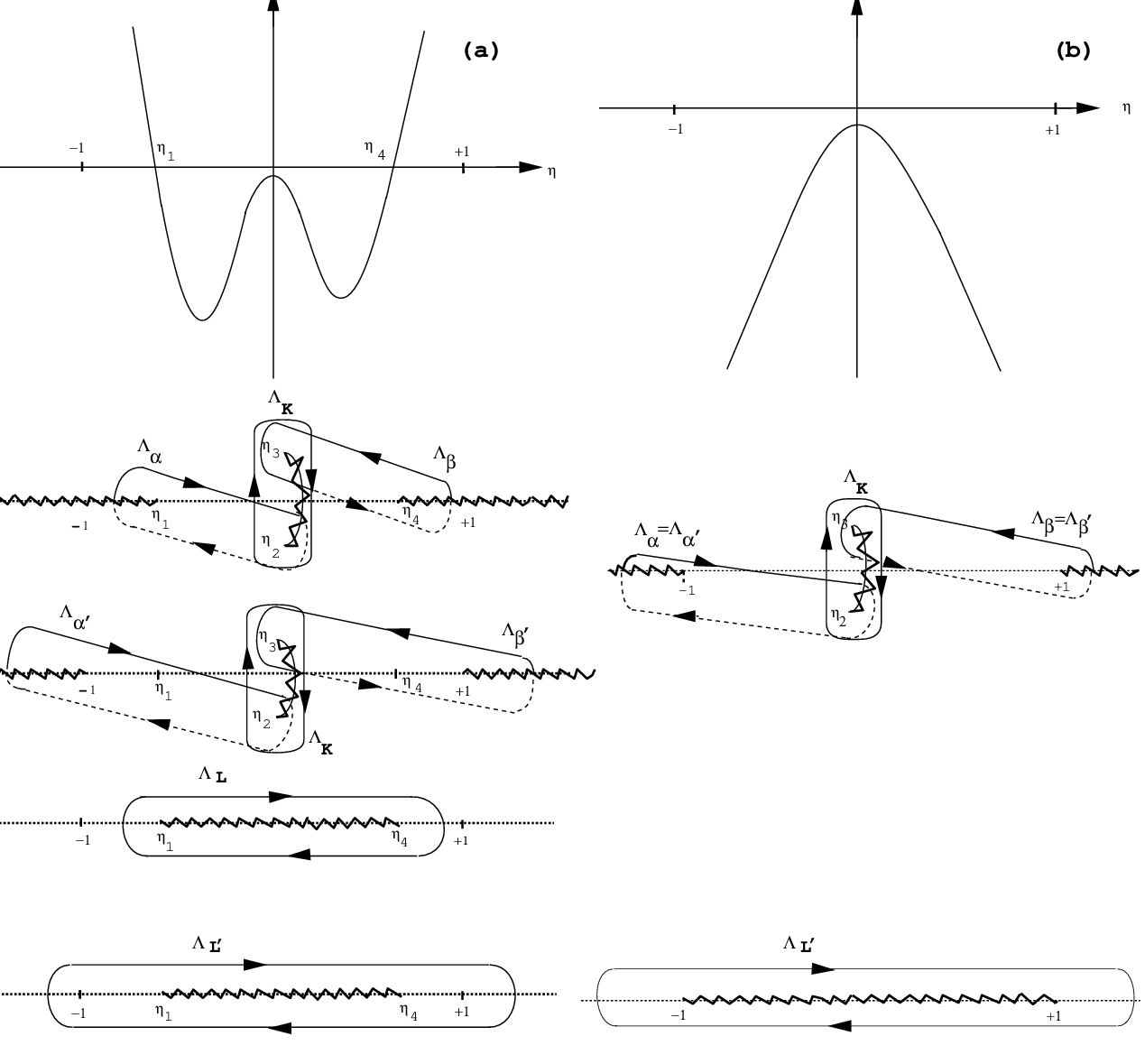, width=\linewidth}}

\caption {The figure gives schematic pictures of $-Q^2(\eta)$ for
$-1<\eta<1$ and of the contours of integration in the complex
$\eta$-plane, when $-Q^2(\eta)$ corresponds to a double-well potential
with an under-dense barrier. In (a) $\Lambda =|m|\ne 0$ and in (b)
$\Lambda =0$.  The cuts are indicated by wavy lines, and the parts of
the contours of integration that lie on  Riemann sheets adjacent to the
complex $\eta$-plane under consideration are dashed. In the left-hand
classically allowed region $Q(\eta)$ is positive. Only those zeros of
$Q^2(\eta)$ that are relevant for the contours of integration are
shown. If the energy lies sufficiently far above the top of the
barrier, one may treat the double-well potential problem as a
single-well potential problem with the classically allowed region
between $\eta_1$ and $\eta_4$ in (a) and between the poles at $\eta=-1$
and $\eta=+1$ in (b). One introduces then new cuts [between $\eta_1$
and $\eta_4$ in (a) and between $-1$ and $+1$ in (b)], on the upper
lips of which $Q(\eta)$ is positive, and uses the contours $\Lambda_L$
and $\Lambda_{L'}$. In (a) the relation between the corresponding
integrals is $L'=L+|m|$. In both (a) and (b) $L$ is related to $\alpha$
and $\beta$ by the relation $L=\alpha+\beta$.}

\label{Fig.4(a)}
\end{figure}

\begin{thebibliography}{10}
\bibitem{ery}
H. Eyring, J.  Walter, and G. E. Kimball, {\it Quantum Chemistry} 
(John Wiley \& Sons and Chapman \& Hall, New York and London, 1944).
\bibitem{her}
G. Herzberg, {\it Molecular Spectra and Molecular Structure}, Volume I. 
{\it Spectra of Diatomic Molecules} (Van Nostrand Reinhold Company, New York,
Cincinnati, Toronto, London, and Melbourne, 1950).
\bibitem{slate}
J. C. Slater, {\it Quantum Theory of Molecules and Solids}, Volume I.
{\it Electronic Structure of Molecules} (McGraw-Hill, New York, 1963).
\bibitem{rose}
B. Rosen, {\it Spectroscopie \'electronique mol\'eculaire}. Encyclopedia 
of Physics Volume {\bf XXVII}, edited by S. Fl\"ugge (Springer-Verlag,
Berlin, G\"ottingen, and Heidelberg, 1964).
\bibitem{bat}
D. R. Bates, K. Ledsham, and A. L. Stewart, Phil. Trans. Roy. Soc.
London {\bf A246}, 215 (1953).
\bibitem{wall}
R. F. Wallis and H. M. Hulburt, J. Chem. Phys. {\bf 22}, 774 (1954).
\bibitem{bat2}
D. R. Bates and T. R. Carson,  Proc. Roy. Soc. {\bf A234}, 207 (1956). 
\bibitem{win}
H. Wind, J. Chem. Phys. {\bf 42}, 2371 (1965). 
\bibitem{peek}
J. M. Peek, J. Chem. Phys. {\bf 43}, 3004 (1965). 
\bibitem{pon}
L. I. Ponomarev and T.P. Puzynina, J. Exp. Theor. Phys. (USSR) {\bf 52},
1273 (1967); English Translation: Sov. Phys. JETP {\bf 25}, 846 (1967).
\bibitem{pon2}
L. I. Ponomarev and  T.P. Puzynina, Joint Institute for Nuclear Research, Dubna,
Preprint P4-3175 (1967). 
\bibitem{hun}
G. Hunter and H.O. Pritchard, J. Chem. Phys. {\bf 46}, 2153 (1967).
\bibitem{wil}
C. M. Rosenthal and E.B.  Wilson, Jr., Int. J. Quant. Chem. {\bf II},
175 (1968). 
\bibitem{bat3}
D. R. Bates and R. H. G. Reid, Adv. in Atom. and Mol. Phys. {\bf 4},
13 (1968).
\bibitem{mur}
T. Murai, Science of Light {\bf 23}, 83 (1974).
\bibitem{mur2}
T. Murai and H. Takatsu,  {\it Tables of Electronic Energy of
$H_2^+$}. Contributions from the Research Group on Atoms and Molecules No. 10,
Progress Report VII, March 1974, p. 74-111. Research Group on Atoms and Molecules,
c/o Department of Physics, Faculty of Science, Ochanomizu University, 1-1,
Otsuko 2 chome, Bunkyo-ku, Tokyo 112, Japan.
\bibitem{wint}
T. G. Winter, M. D. Duncan, and N. F. Lane, J. Phys. B: Atom. Molec.
Phys. {\bf 10}, 285 (1977).
\bibitem{kla}
M. Klaus, J. Phys. A: Math. Gen. {\bf 16}, 2709 (1983). 
\bibitem{bye}
W. Byers Brown and E. Steiner, J. Chem. Phys. {\bf 44}, 3934 (1966).
\bibitem{strand}
M. P. Strand and W.P. Reinhardt, J. Chem. Phys. {\bf 70}, 3812 (1979).
\bibitem{paj}
P. Pajunen, Mol. Phys. {\bf 43}, 753 (1981).
\bibitem{fro10}
N. Fr\"{o}man and P.O. Fr\"{o}man, {\it PHASE-INTEGRAL METHOD
Allowing Nearlying Transition Points}, With adjoined papers by A. Dzieciol,
N. Fr\"oman, P.O. Fr\"oman, A. H\"okback. S. Linnaeus, B. Lundborg, and E. Walles.
Springer Tracts in Natural Philosophy Vol. 40, edited by C. Truesdell
(Springer-Verlag, New York, Berlin, and Heidelberg, 1996).
\bibitem{lak}
M. Lakshmanan and P. Kaliappan, J. Phys. A: Math. Gen. {\bf 13}, L299 (1980).
\bibitem{lak2}
M. Lakshmanan, F. Karlsson, and P. O. Fr\"{o}man, Phys. Rev. {\bf D24}, 2586 (1981).
\bibitem{lak3}
M. Lakshmanan, P. Kaliappan, K. Larsson, F. Karlsson, and P. O. Fr\"{o}man, Phys. 
Rev. {\bf A49}, 3296 (1994).
\bibitem{dam}
\"{O}. Dammert and P. O. Fr\"{o}man, J. Math. Phys. {\bf 21}, 1683
(1980).
\bibitem{fro8}
N. Fr\"{o}man and P. O. Fr\"{o}man, {\it JWKB Approximation,
Contributions to the Theory} (North-Holland, Amsterdam, 1965. 
Russian translation: MIR, Moscow, 1967).
\bibitem{fro16}
N. Fr\"{o}man, P. O. Fr\"{o}man, U. Myhrman, and R. Paulsson,
Ann. Phys. (NY) {\bf 74}, 314 (1972).
\bibitem{car}
N. Fr\"oman and P. O. Fr\"oman, {\it On the History of the so-called WKB-Method
from 1817 to 1926}. Proceedings of the Niels Bohr Centennial Conference,
Copenhagen 25-28 March 1985 on {\it Semiclassical Descriptions of Atomic
and Nuclear Collisions}, p. 1-7, edited by Jens Bang and Jorrit de Boer 
(North-Holland, Amsterdam, Oxford, New York, and Tokyo, 1985).
\bibitem{fro13}
N. Fr\"{o}man, P. O. Fr\"{o}man, and K. Larsson, Phil. Trans. Roy.
Soc. London {\bf A347}, 1 (1994).
\bibitem{fro14}
N. Fr\"{o}man, P. O. Fr\"{o}man, and B. Lundborg, Math. Proc.
Camb. Phil. Soc. {\bf 104}, 153 (1988).
\bibitem{fro5}
N. Fr\"{o}man, Ann. Phys. (NY) {\bf 61}, 451 (1970).
\bibitem{fro11}
N. Fr\"oman and W. Mrazek, J. Phys. A: Math. Gen. {\bf 10}, 1287
(1977).
\bibitem{fro4a}
N. Fr\"{o}man, Ark. Fys. {\bf 31}, 381 (1966).
\bibitem{fro9}
N. Fr\"{o}man and P. O. Fr\"{o}man, Nucl. Phys. {\bf A147}, 606
(1970).
\bibitem{fro7}
N. Fr\"{o}man, {\it Semiclassical and Higher-Order Approximations.
Properites. Solution of Connection Problems.} Article in: {\it
Semiclassical Methods in Molecular Scattering and Spectroscopy,}
edited by M. S. Child (Reidel, Dordrecht, Boston, and London, 1980).
\bibitem{child}
M. S. Child, {\it Molecular Collision Theory} (Academic Press, London
and New York, 1974). Reprinted with corrections in 1984.
\bibitem{fro15}
N. Fr\"{o}man, P. O. Fr\"{o}man, and B. Lundborg,  Adjoined paper
(Chapter 5) in \cite{fro10}.
\bibitem{fro4c}
N. Fr\"{o}man, Ark. Fys. {\bf 32}, 541 (1966).
\bibitem{fro6}
N. Fr\"{o}man, Phys. Rev. {\bf A17}, 493 (1978).
\bibitem{paul}
R. Paulsson, F. Karlsson, and R. J. Leroy, J. Chem. Phys. {\bf 79}, 4346 (1983).
\bibitem{fro4b}
N. Fr\"{o}man, Ark. Fys. {\bf 32}, 79 (1966).
\bibitem{fro12}
N. Fr\"oman and U. Myhrman, Ark. Fys. {\bf 40}, 497 (1970).

\end{thebibliography}
\end{document}